\documentclass{aastex62}

  \newcommand{\bxi}{{\boldsymbol\xi}}
  \newcommand{\bx}{{\boldsymbol x}}
  \newcommand{\noise}{{\mathcal N}}
  \newcommand{\target}{{\mathcal T}}

  \usepackage{amsmath}
  \usepackage{graphicx}
  \graphicspath{{./}{figures/}}

  \submitjournal{ApJL}

  \shorttitle{Rossby waves in the Sun}
  \shortauthors{Hanasoge \& Mandal}

\begin{document}

  \title{Detection of Rossby waves in the Sun using normal-mode coupling}

  \correspondingauthor{Shravan Hanasoge}
  \email{hanasoge@tifr.res.in}

  \author[0000-0002-0786-7307]{Shravan Hanasoge}
  \affil{Department of Astronomy \& Astrophysics,
  Tata Institute of Fundamental Research, Mumbai 400005}
  \affiliation{Center for Space Science, New York University Abu Dhabi, UAE}
  \author{Krishnendu Mandal}
  \affil{Department of Astronomy \& Astrophysics,
  Tata Institute of Fundamental Research, Mumbai 400005}

  \begin{abstract}

  Rossby waves play a fundamental role in angular momentum processes in rotating
  fluids. In addition to the potential to shed light on physical mechanisms operating in the solar convection zone, the recent detection of Rossby waves in the Sun \citep{gizon18,liang_2018} also
   serves as a means of comparison between different helioseismic methods.
  Time-distance helioseismology, 
  ring-diagram analysis and other techniques have all proven
  successful in recovering the Rossby-wave dispersion relation from analyses of Helioseismic and 
  Magnetic Imager data \citep[HMI;][]{hmi}. In this article, we demonstrate that analyses of two years of HMI global-mode-oscillation data using the technique of normal-mode coupling also show signatures of Rossby waves. In addition to providing an independent means of inferring Rossby waves, this detection lends credence to the methodology of mode coupling and encourages a more complete exploration of its possibilities. 

  \end{abstract}



  \section{Introduction} \label{sec:intro}
 Rossby waves have been a subject of interest in solar physics, since theoretical arguments \citep{Papa1978,saio_82,provost_81,Wolff_1986, unno_1989} 
strongly favor their existence {for a rotating spherical fluid body such as the Sun}, but their detection has proven elusive {until recently. \citet{Kuhn2000} found uniformly spaced 100 $m$ high ``hills" on the solar photosphere and ascribed it to Rossby waves. \citet{ulrich2001} found evidence of very long-lived wave patterns on the Sun with azimuthal number $m\le 8$. \citet{sturrock2015} claimed to have found signatures of Rossby waves with azimuthal number $m=1$ from measurements of changes in solar radius. \citet{gizon18} applied granulation-tracking and ring diagram analysis \citep{hill88} to 6 years of HMI data and discovered evidence of a Rossby-wave dispersion relation. {\citet{liang_2018} also confirmed the detection by analyzing $21$ years of space-based data using time-distance helioseismology \citep{Duvall_1993}.} They were only able to detect sectoral Rossby modes, i.e. equatorially focused retrograde-propagating features with $t=-s$, where $t$ is azimuthal order and $s$ is spherical-harmonic degree, suggesting that latitudinal differential rotation likely filters out the other modes \citep{yoshida2000,Wolff1998}. Rossby waves are of particular interest to helioseismology because they are non-axisymmetric, time-varying and robust, appearing with distinct clarity in different seismic analyses. 

  The technique of normal-mode coupling has a long and illustrious history \citep[e.g.][]{DT98}. Mode coupling was first introduced in terrestrial seismology, where it continues to be used to infer the non-axisymmetric, anisotropic structure of the Earth. Coupling theory applied to the Sun, often equivalently termed ``eigenfunction distortion" or ``eigenfunction mixing" for reasons that will become apparent shortly, was first described in its modern form by \citet{woodard_89}. While the technique has subsequently seen increasing purchase, e.g. \citet{lavely92,roth_stix_2008,woodard07,vorontsov11,woodard14,woodard16,schad,schad_2013,hanasoge17_etal,hanasoge17,hanasoge18}, it has yet to enter the mainstream and become well adopted. A significant obstacle is the related mathematical complexity and the abstracted nature of the measurements (a sequence of complex numbers, for instance), hindering straightforward interpretation. However, the power of the method lies in the remarkable simplicity of the data-handling procedure \citep{woodard16, hanasoge18}, especially compared with other helioseismic techniques used for inferring non-axisymmetric variations in the Sun such as time-distance \citep{duvall}, ring-diagram analysis \citep{hill88} or acoustic holography \citep{lindsey97}. The balance between arduous mathematics associated with interpretation and the simplified nature of the measurement is appealing because of the limited extent of data massaging required. An important aspect to note is that the technique described by \citet{woodard16} and \citet{hanasoge17_etal} directly takes into account time variability, thereby allowing for the inference of solar internal structure and flows as a function of spatial wavenumber and temporal frequency. 

  Mode coupling is based on the idea that the linear (symmetric / self-adjoint) wave operator of the Sun has an orthonormal and complete basis of eigenfunctions $\bxi_k(\bx)$, where $\bx$ is space and $k$ is the  associated index, which in spherical geometry typically consists of 3 quantum numbers spherical-harmonic degree $\ell$, azimuthal order $m$ and radial order $n$ \citep[e.g.][]{goedbloed2004, jcd_notes}. The wavefield $\bxi$ is a weighted sum of the eigenfunctions, given by 
\begin{equation}
\bxi = \sum_{k} a^\omega_k\, \bxi_k,
\end{equation}  
where $\omega$ is temporal frequency and $a^\omega_k = a_k(\omega)$ encodes power-spectral information such as the Lorentzian, phase etc. Orthonormality implies that the following relationship holds for both solar and reference-model eigenfunctions,
\begin{equation}
\int_\odot d\bx\,\bxi^*_{k'}\cdot\bxi_k = \delta_{kk'}.
\end{equation}
Many aspects of complex physics such as convection and magnetism which the Sun exhibits are not present in our models of solar structure \citep[such as model S;][]{jcd}, and therefore the eigenfunctions of the Sun will necessarily be different than those of the reference model. However, completeness and orthonormality allow us to express the eigenfunctions of the Sun as a unique, linear weighted sum of those of the reference model,
\begin{equation}
\bxi^{\odot}_k = \sum_{k{'}} c^{k{'}}_k\,\bxi^{\rm ref}_{k{'}}.\label{coupeq}
\end{equation}
The weights $c^{k{'}}_k$, known as coupling coefficients, may be directly measured from observations \citep{woodard16,hanasoge18} of the wavefield at the solar surface $\phi^\omega_{k}$,i.e. $c^{k{'}}_k \propto \int d\omega\,W^{\omega}_{kk'}\,\phi^{\omega*}_k\,\phi^{\omega+\sigma}_{k'}$,
where the spatial scale associated with the perturbation is related to the distance between the mode wavenumbers $k$ and $k'$ and the temporal evolution to the frequency offset $\sigma$. The term $W^\omega$ is a weight function which may be derived based on theoretical considerations. The implication of equation~(\ref{coupeq}) is that we compute the coupling coefficients at each frequency channel $\sigma$, giving us access to the spatio-temporal structure of the perturbation. Through tedious algebra \citep{lavely92,hanasoge17_etal,hanasoge18}, these coupling coefficients may be related to the difference between properties of the Sun and the reference model. It is important to note that the theory only functions in the limit of linear perturbation theory, i.e. for suitably small deviations and where the temporal scale of of the perturbation is much longer than mode periods, i.e. $\sigma \ll \omega_{k}$.

  In this Letter, we describe the first detection of non-axisymmetric, time-varying Rossby waves using normal-mode coupling, adding to the preponderance of evidence and validating the method. Since we do not observe the entire surface of the Sun,  i.e. we are only seeing effectively a third, spatial windowing induces spectral broadening. As a consequence, different spatial harmonics leak into each other and we are unable to perfectly isolate individual spherical harmonics. This effect may be modeled in detail \citep{hanasoge18} using carefully computed leakage matrices \citep{schou94}, but for the present analysis, we only retain diagonal terms, i.e. the amplitude of a specific wavenumber after accounting for leakage.

  \section{Data analysis}
  We use 2 years of the global mode time series' data in the mode range $\ell \in [50,170]$, where $\ell$ is spherical-harmonic degree, taken by the Helioseismic and Magnetic Imager \citep[HMI; available for download from the Stanford data repository, http://jsoc.stanford.edu/ ;][]{hmi}. To improve the signal-to-noise ratio and frequency resolution, we analyze the entire dataset at one go. The measurement is $\phi^{\omega*}_{\ell m}\,\phi^{\omega+\sigma}_{\ell m+t}$, where $\phi^\omega_{\ell m} = \phi_{\ell m}(\omega)$ is the temporal Fourier transform of the global-mode time series $\phi$ for mode $(\ell, m)$ and $m$ is azimuthal order. Temporal frequency is denoted by $\omega$ and the offset $\sigma$ represents the time-scale associated with the perturbation. Because we are correlating the wavefield at different azimuthal orders, the measurement is sensitive to non-axisymmetric features with azimuthal order $t$ and harmonic degree $s$. In order to deal with more compact measurements, we calculate $B$-coefficients, which are linear-least-square fits to the raw wavefield correlations, defined thus
  \begin{equation}
  B^\sigma_{st} = \frac{\sum_{m\omega} H^\sigma_{\ell m s t}(\omega)\phi^{\omega*}_{\ell m}\,\phi^{\omega+\sigma + t\Omega}_{\ell m+t}}{\sum_{m\omega} |H^\sigma_{\ell m s t}|^2},
  \end{equation}
  where $H$ is a function comprising mode-normalization constants, diagonal leakage elements \citep{schou94}, Wigner-3$j$ symbols and the power-spectral model \citep[for details on how $H$ is evaluated, see][]{woodard16,hanasoge17_etal,hanasoge18}, 
\begin{equation}
H^\sigma_{\ell m s t}(\omega) = -2\omega(-1)^{m+t}\sqrt{2s+1}\begin{pmatrix} \ell & s & \ell \\ -(m+t) & t & m\end{pmatrix}
L_{\ell m}^{\ell m}L_{\ell	m+t}^{\ell	m+t}N_\ell(R^{\omega*}_{\ell m} | R^{\omega+\sigma+t\Omega}_{\ell m+t}|^2 +
|R^{\omega}_{\ell m}|^2  R^{\omega+\sigma+t\Omega}_{\ell m+t} ),\label{eqH}
\end{equation}
where the first term on the right-hand side of equation~(\ref{eqH}) is a Wigner-3$j$ symbol and $L^{\ell' m'}_{\ell m}$ is the leakage from mode $(\ell,m)$ to $(\ell',m')$. Only diagonal leakage terms $L^{\ell m}_{\ell m}$ are retained in this analysis. The power spectrum of a mode can be described by a Lorentzian \citep{anderson_1990,Duvall_harvey_93},
\begin{equation}
R^\omega_{\ell m} = \frac{1}{(\omega_{n\ell m}-i\Gamma_{n\ell}/2)^2 - \omega^2},
\end{equation}
where {$\omega_{n\ell m}$ is the resonant frequency of the mode (with radial order $n$) and $\Gamma_{n\ell}$ is its inverse lifetime. }
  The rotation of the Sun advects features along with and we therefore track the data at a suitable rate - this results in frequency shifting each azimuthal order by $t\Omega$. The Sun is differentially rotating, implying that a variety of rotation rates $\Omega$ may be chosen in order to track the data. Here, we fix it at $\Omega = 453$ nHz, the rate corresponding to the equatorial rotation. With respect to this co-rotating frame, $t>0$ and $t<0$ correspond to prograde and retrograde propagating features respectively.
  Not taking into leakage \citep[this complicates the analysis significantly;][]{hanasoge18}, the $B$ coefficients are directly sensitive to the properties of the medium \citep{woodard14,hanasoge18}, 
  \begin{equation}
  B^\sigma_{st}(n,\ell) = \sum_{s}\int_\odot dr\, w^\sigma_{st}(r)\,f_{s} K_{n\ell}(r),\label{inverse}
  \end{equation}
  where $K_{n\ell}(r)$ is the sensitivity kernel comprising the eigenfunction associated with mode $(n,\ell)$ and $f_s$ is a term obtained from asymptotic analysis of the kernels \citep{vorontsov11,woodard14,hanasoge18}. {$f_s$ is only non-zero for odd $s$ \citep[see e.g. Equation (6) of][]{hanasoge18}, i.e., we are only able to infer Rossby modes with odd harmonic degrees using measurements of coupling between oscillations of the same harmonic-degree $\ell$. Measuring even-$s$ Rossby modes requires the analysis of coupling of oscillations of different harmonic degrees, a greater challenge and one that is reserved for future work.} $B$ coefficients computed for self-coupled modes are only sensitive to the toroidal-flow component, which is given by 
  \begin{equation}
  {\bf u}^\sigma(r,\theta,\phi) = \sum_{st} w^\sigma_{st}(r)\,{\bf e}_r\times {\boldsymbol\nabla}_h Y_{st}(\theta, \phi),
  \end{equation}
  where $(r,\theta,\phi)$ and $({\bf e}_r,{\bf e}_\theta,{\bf e}_\phi)$ are radius, co-latitude and longitude and respective unit vectors, ${\boldsymbol\nabla}_h = {\bf e}_\theta\partial_\theta + {\bf e}_\phi (\sin\theta)^{-1}\partial_\phi$ and $Y_{st}$ is the spherical harmonic associated with harmonic-degree $s$ and azimuthal-order $t$. Toroidal flow is, by construction, mass conserving and does not possess a radial component, indicating that it is directly equal to the radial vorticity times a factor of $r$.	

  We compute $B$-coefficients for all identified radial orders associated with harmonic degrees in the range $\ell \in [50,170]$. Since we are interested in capturing time-varying phenomena, we study a finite range in $\sigma \in [0, 0.5]\, \mu$Hz, allowing us to explore low-frequency non-axisymmetric evolution of toroidal flows. For the present analysis, we limit the harmonic degree $s \le 20$, since the Rossby-wave frequency rapidly drops with increasing $s$. 

\section{Inversion and Rossby waves}
The toroidal-flow coefficients $w^\sigma_{st}$ are linearly related to the measured $B$-coefficients through the integral~(\ref{inverse}), allowing us to pose an inverse problem connecting the two. In this analysis, we apply the method of Subtractive Optimally Localized Averaging \citep[SOLA][]{pijpers} to determine the coefficients $\alpha^\sigma_{n\ell;r_0}$ such that the toroidal flow at a specific depth $r_0$ is obtained through a weighted average of the $B$ coefficients,
\begin{equation}
w^\sigma_{st}(r_0) = \sum_{n\ell}\alpha^\sigma_{n\ell;r_0}\,B^\sigma_{st}(n,\ell).
\end{equation}
Spatial windowing in the observed data, which is a rotating, non-axisymmetric, temporally varying system, implies that leakage likely occurs both spatially and temporally (such as displaying spurious higher-frequency harmonics). Additionally, leakage limits the ability to isolate azimuthal order $t$, indicating that the frequency offset due to tracking, $t\Omega$, may also introduce errors. A more detailed analysis, which involves optimizing each spatio-temporal bin \citep{hanasoge18}, is required to account for all these issues. For the present analysis, we ignore both these effects and assume $\alpha^\sigma_{n\ell;r_0} \approx \alpha_{n\ell;r_0}$. We determine $\alpha^\sigma_{n\ell;r_0}$ by optimizing the cost functional
\begin{equation}
\chi = \int_\odot dr\,[\target(r;r_0) - \sum_{n\ell} \alpha_{n\ell;r_0} f_s K_{n\ell}(r)]^2 + \lambda \sum_{n\ell} \noise_{n\ell} \,\alpha^2_{n\ell;r_0},\label{chi}
\end{equation}
where $\target(r;r_0)$ is a desired target function, such as a Gaussian in radius centered around $r_0$, $\noise_{n\ell}$ is the diagonal component of the noise-covariance matrix and $\lambda$ is a regularization parameter. Evaluating the noise matrix involves tedious algebra, a detailed description of which may be found in \citet{hanasoge18}. We determine $\lambda$ by identifying the knee of the curve that plots the misfit between the averaging kernel $\sum_{n\ell} \alpha_{n\ell;r_0}   f_s K_{n\ell}(r)$ and the target against the noise. Here, we only perform inversions for relatively shallow layers $r/R_\odot = 0.97, 0.99$, leaving a more detailed inversion as a function of depth for future work. {In Figure \ref{fig:avg_kernel}, we show the L curve obtained by varying the regularization parameter $\lambda$ for the wavenumber $(s,t)=(11,-11)$ at a depth of $r/R_\odot = 0.99$ and {corresponding averaging kernel}. The knee of the L curve represents the optimum tradeoff - i.e., it is the best possible fit to the target associated the lowest noise level (right panel of Figure~\ref{fig:avg_kernel}).}

Rossby waves propagate in a rotating spherical fluid ball according to the dispersion relation $\omega_R = -2\Omega\,t/[s(s+1)]$ \citep{saio_82}, where $\omega_R$ is the frequency of the Rossby wave and $\Omega$ is the rotation rate. Physically, Rossby wave solutions are solely retrograde propagating. {In Figure \ref{fig:colorplot}, we plot $\vert w_{st}\vert^2$ at two different depths $0.99 R_\odot$ and $0.97 R_\odot$, obtained from the inversion of the measured $B$ coefficients by using Equation (\ref{inverse}). It can be seen that there is significant power close to the theoretically predicted frequencies of retrograde-propagating sectoral Rossby waves, $\omega_R = 2\Omega/(s+1)$, with $\Omega = 453$ nHz. Normalized power for different harmonic degrees $s$ is shown in Figure \ref{fig:cut_plot}. Note that the theoretically estimated value of the frequency of the modes in a uniformly rotating frame might be different from true value because of solar differential rotation \citep{Wolff1998} and other complex effects, e.g magnetic field and convection. 

The detection of the $s=1$ mode as proposed by \citet{sturrock2015}, which \citet{gizon18,liang_2018} are unable to find because of the limitation of their measurement technique, must be treated with caution. Since the tracking rate is same as the frequency of this particular mode, systematics in our technique might induce spurious power in that spatial and frequency bin. We find signature of sectoral modes for odd harmonic degree as high as $s\approx15$ in this analysis. This is in line with the theoretical suggestion by \citet{Wolff1998} and the analyses of \citet{gizon18,liang_2018}, who also reached the same conclusion. Rossby waves in the Sun appear to be equatorially confined retrograde {\it sectoral} modes, perhaps attributed to latitudinal differential rotation \citep{Wolff1998} which likely filters out tesseral and zonal-like Rossby modes, i.e. solutions that seep into high latitudes. }

\begin{figure}
\begin{centering}
\includegraphics[scale=0.45]{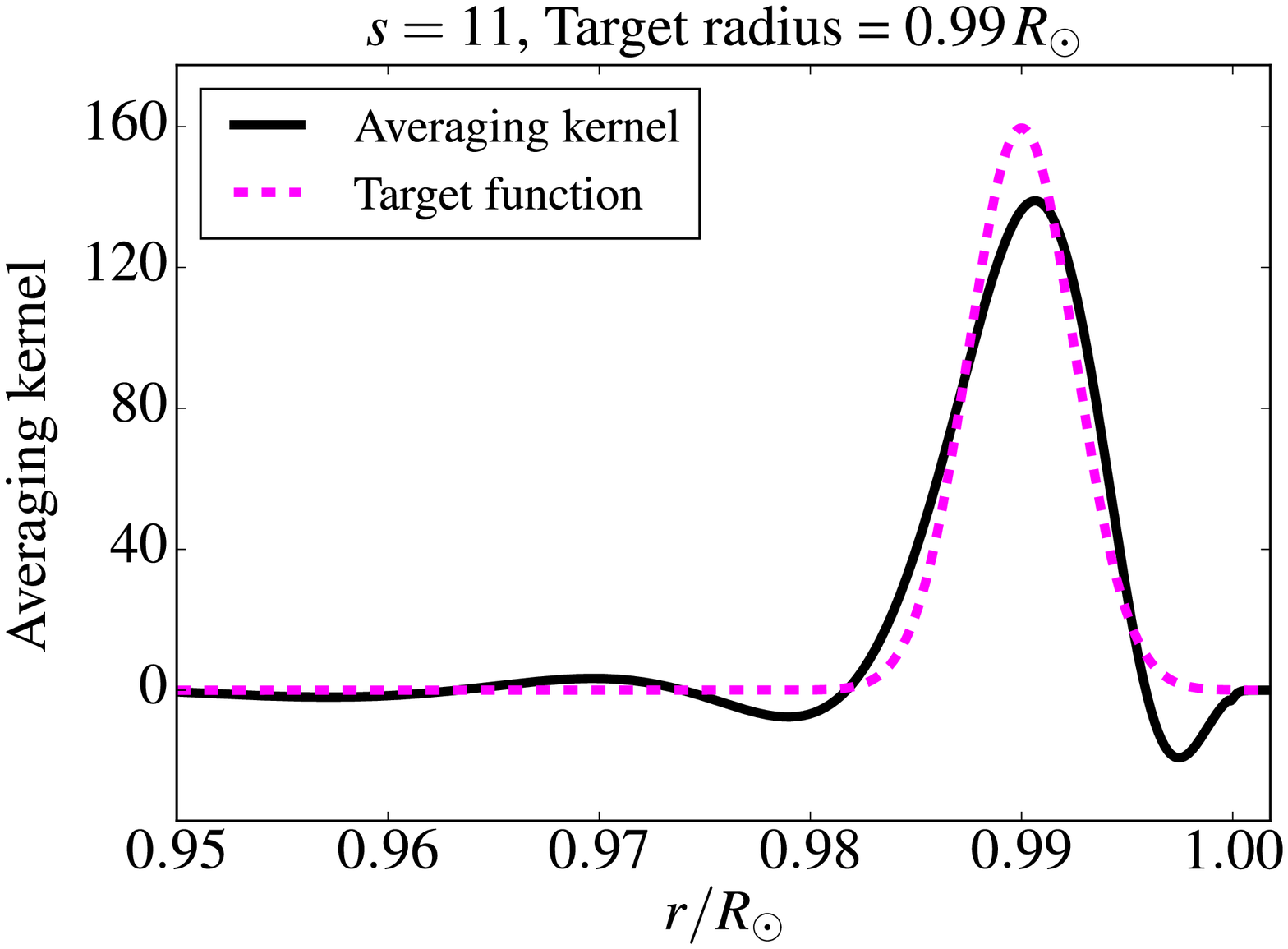}\includegraphics[scale=0.45]{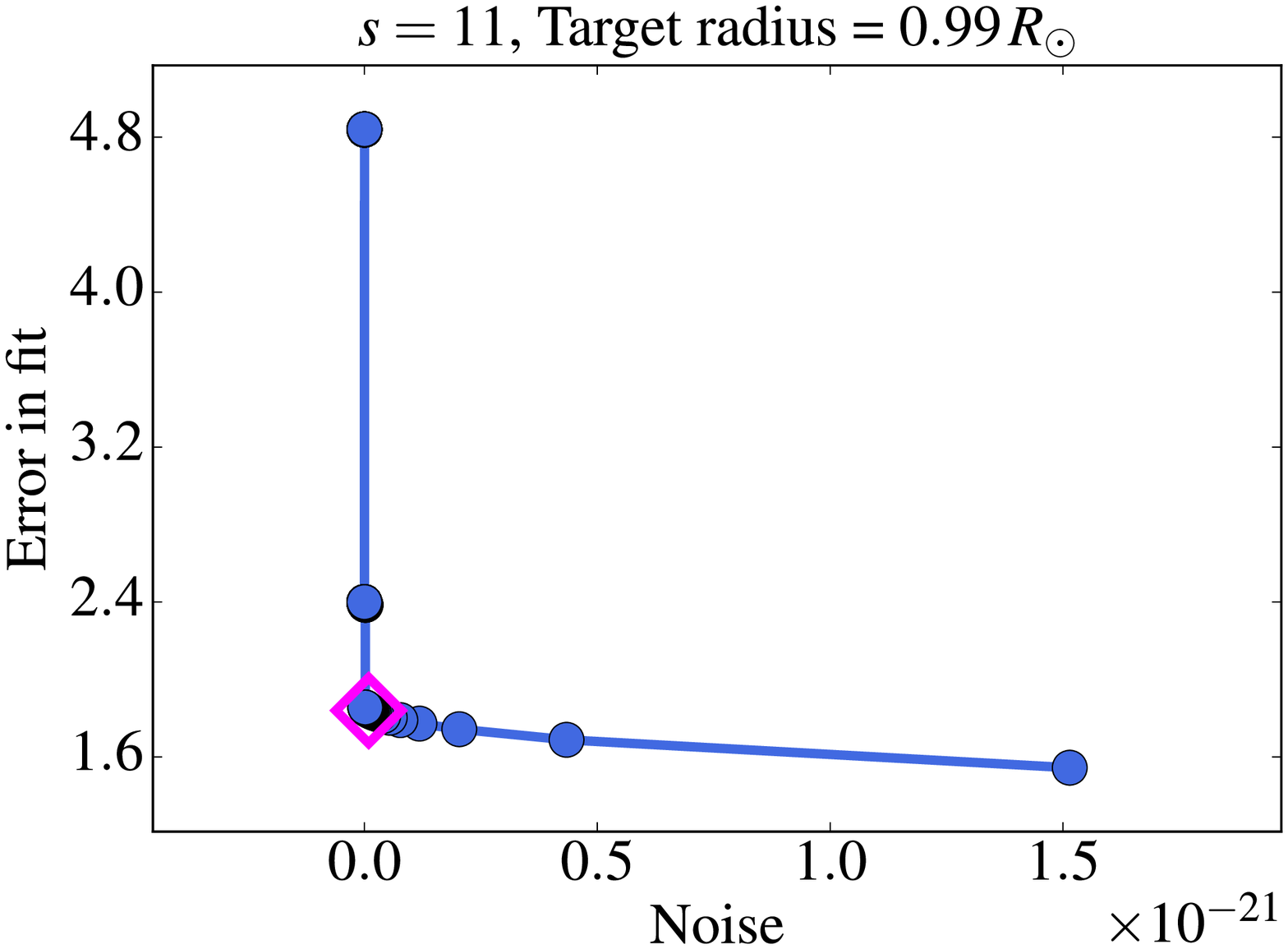} 
\caption{\label{fig:avg_kernel}{Left panel: black-solid line and red-dashed line show averaging kernels for harmonic degree $(s,t)=(11,-11)$ and corresponding target function at a depth of $0.99R_\odot$ respectively. Right panel: Tradeoff L curve (blue-solid line with blue circles) is plotted for the same wavenumber and same depth as used for the plot in the left panel by varying the smoothing parameter $\lambda$. The $y$ axis of the plot shows the error in fitting the desired target function $\target(r;r_0)$ in Equation-(\ref{chi}) and the $x$-axis shows the amplification of the noise from inversion, which is the second term of Equation-(\ref{chi}).  The value of $\lambda$ associated with the knee of the L curve, marked by the red-diamond marker, is chosen to be the regularization parameter.}}
\par\end{centering}
\end{figure}

\begin{figure}
\begin{centering}
\includegraphics[scale=0.45]{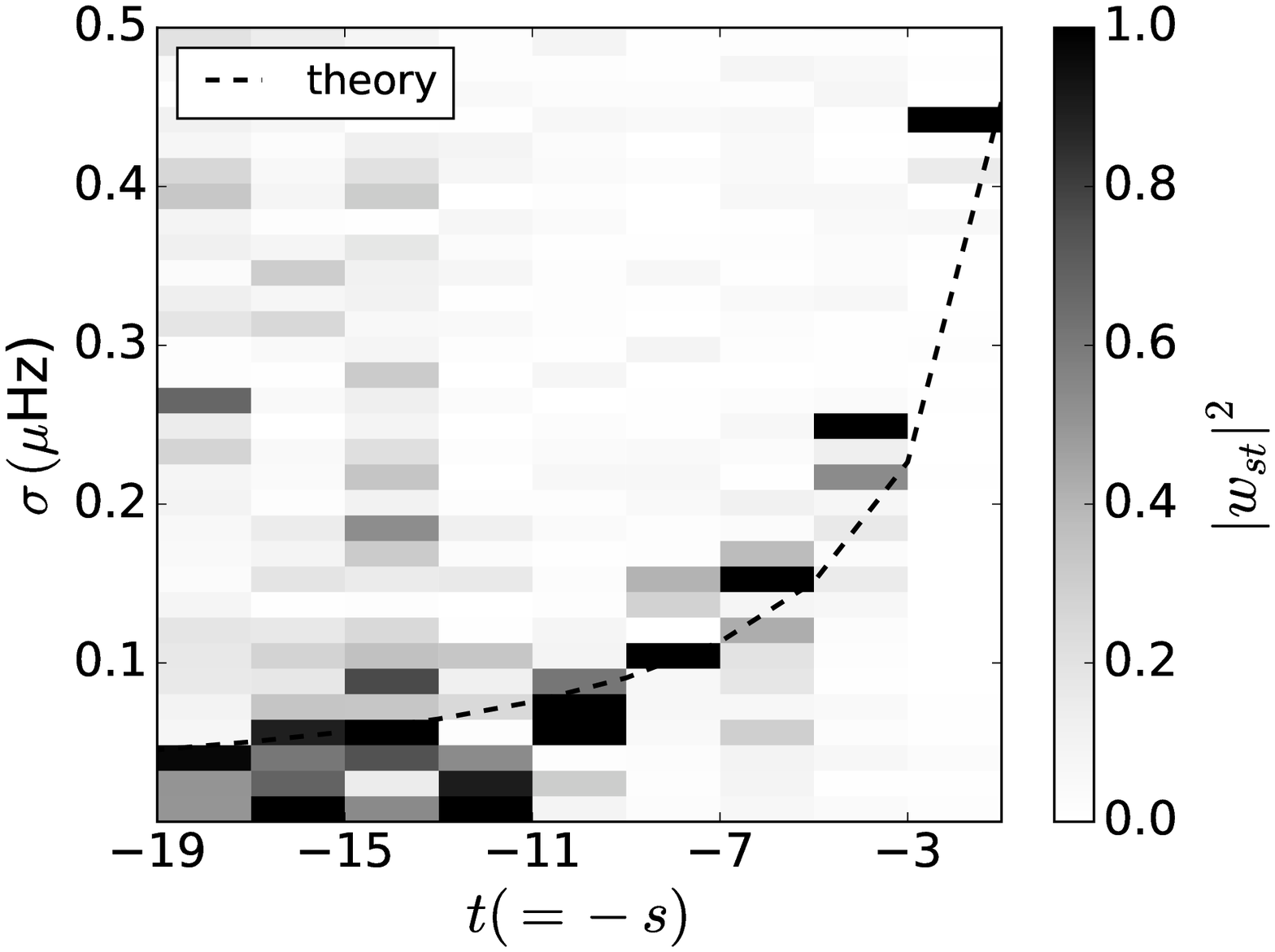}\includegraphics[scale=0.45]{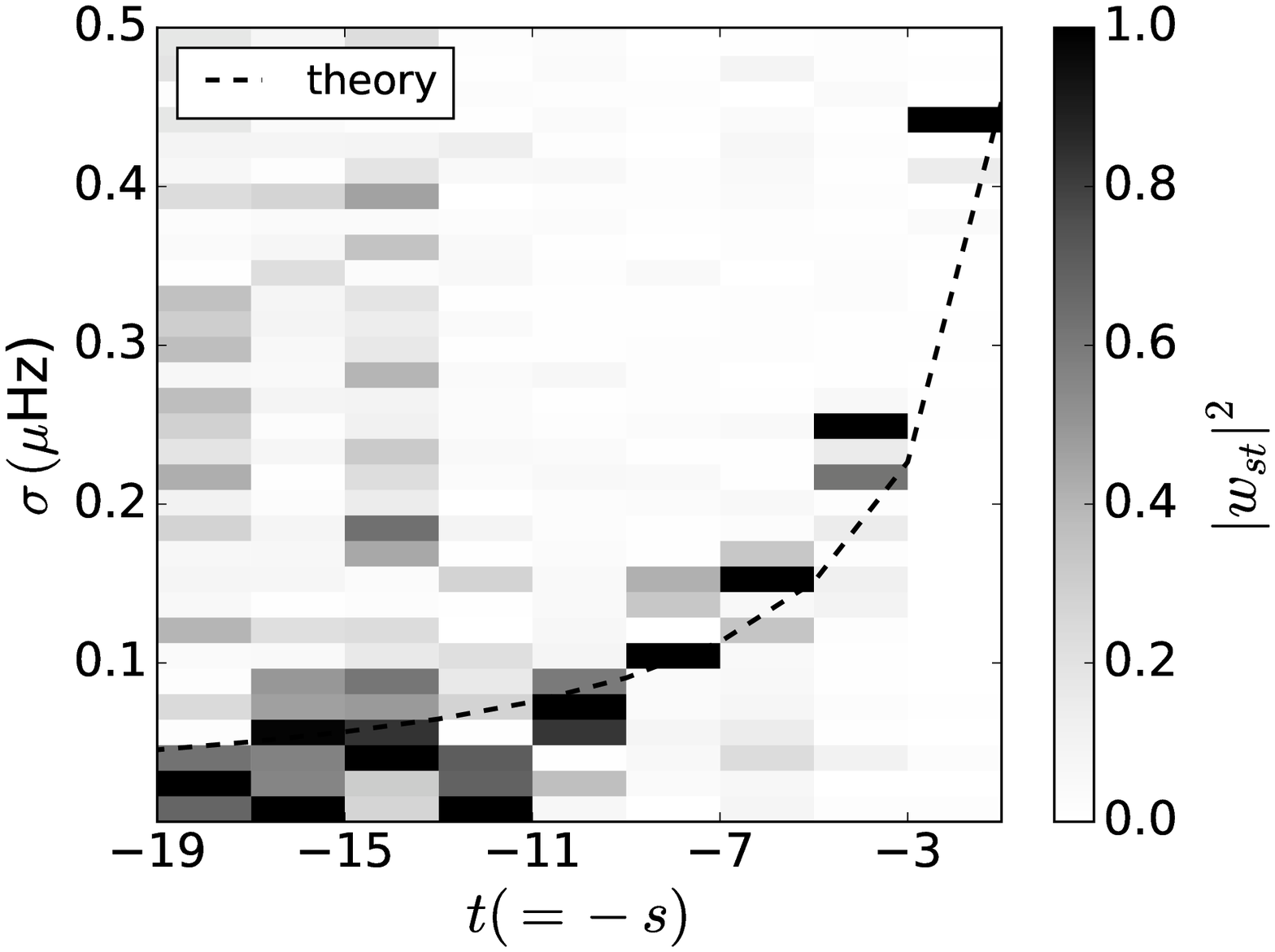} 
\caption{\label{fig:colorplot}{Normalized power spectrum of Rossby modes obtained by inverting the $B$ coefficients. Only modes with odd $s$, starting from $s=1$ can be obtained using our measurement technique. The left and right panels show inversion results for depth $0.99R_\odot$ and $0.97\,R_\odot$, respectively. The black dashed line corresponds to the theoretical dispersion relation of sectoral Rossby modes $\omega_R = 2\Omega/(s+1)$ in a uniformly rotating frame with rotation $\Omega = 453$ nHz.}}
\par\end{centering}
\end{figure}

\begin{figure}
\begin{centering}
\includegraphics[scale=0.5]{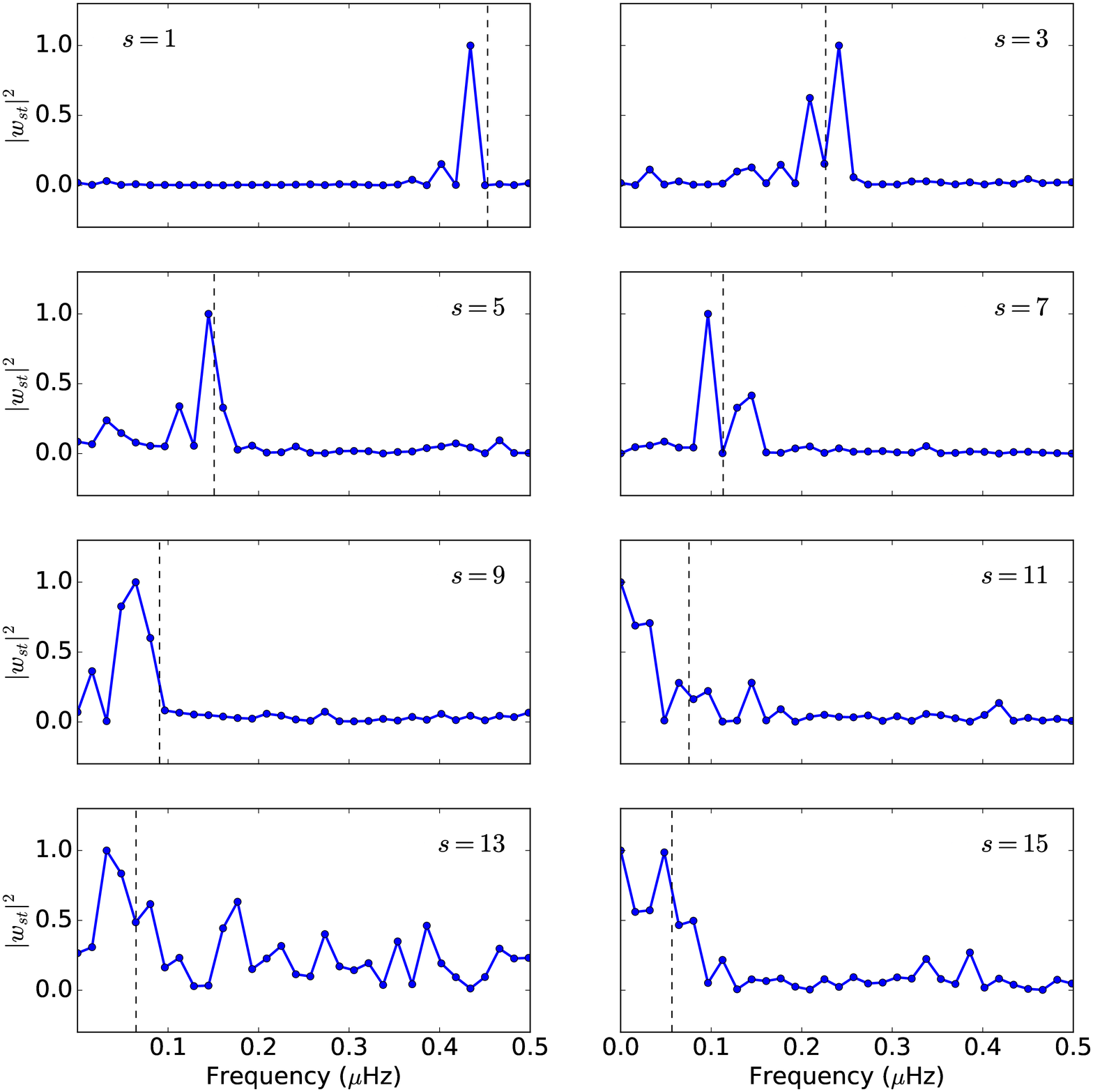}
\caption{\label{fig:cut_plot}{Normalized power spectra of sectoral modes of Rossby waves with odd harmonic degree, $s$ (the values of which are mentioned in each panel) at depth $0.97R_\odot$. The black vertical dashed line in each panel shows the theoretical value of the frequency for that particular retrograde-propagating mode $\omega_R = 2\Omega/(s+1)$, calculated in a rotating frame with rotation $\Omega = 453$ nHz.}}
\par\end{centering}
\end{figure}

\section{Conclusions}
The detection of Rossby waves represents an important milestone for normal-mode coupling, serving to verify that the measurement is able to accurately capture spatial non-axisymmetry and temporal evolution. It will be extremely valuable to perform detailed comparisons between inferences of Rossby waves using different helioseismic techniques.  {In this work, we report the detection of sectoral modes of Rossby waves with odd harmonic degree, starting from $s=1$ up to $s\approx 15$. In a future, more detailed analysis, we will categorize the depth dependence of Rossby waves, their lifetimes and their evolution over the solar cycle.} The impact of instrumental systematics, spatial leakage and noise modeling will be fully incorporated. 

\acknowledgments
 SMH acknowledges support from the Ramanujan fellowship SB/S2/RJN-73/2013 and the Max-Planck partner group program. {K. M acknowledges financial support from Department of Atomic Energy, India.}

  \bibliography{references}
  \end{document}